\documentclass[12pt]{article}
\usepackage[margin=1.5in]{geometry}

\usepackage{times}
\usepackage{graphicx}
\usepackage[font=small]{caption}
\usepackage{subcaption}
\usepackage{float}
\usepackage{bm}
\usepackage[square, numbers, sort&compress]{natbib}
\usepackage{url}
\usepackage{amsmath}
\usepackage{hyperref}
\usepackage{microtype} 
\newcommand{\be}{\begin{equation}}
\newcommand{\ee}{\end{equation}}
\newcommand{\bea}{\begin{eqnarray}}
\newcommand{\eea}{\end{eqnarray}}

\usepackage{titlesec}
\titleformat{\section}
  {\normalfont\normalsize}
  {\thesection}{1em}{}
\titleformat{\subsection}
  {\normalfont\small}
  {\thesubsection}{1em}{}

\title{\bf \large Potential Pitfalls in Visual Models of Tipping Points - And How to Fix Them}
\author{\normalsize Jonathan Dechert$^{1*}$, Svetlana Gurevich$^2$ and Stefan Heusler$^1$\\
\small$^1$ Institut f\"ur Didaktik der Physik, Universit\"at M\"unster, Germany\\
\small$^2$ Institut f\"ur Theoretische Physik, Universit\"at M\"unster, Germany\\
\small$^*$ jdechert@uni-muenster.de}

\date{}

\begin{document}

\maketitle

\begingroup
\renewcommand{\thefootnote}{} 
\footnotetext{Author accepted manuscript. Published in \emph{Northeast Journal of Complex Systems (NEJCS)}, Vol. 7, No. 3, Article 3 (2025). DOI: \href{https://doi.org/10.63562/2577-8439.1113}{10.63562/2577-8439.1113}. Available at: \url{https://orb.binghamton.edu/nejcs/vol7/iss3/3}.}
\addtocounter{footnote}{-1} 
\endgroup

\thispagestyle{empty}

\enlargethispage{3\baselineskip}

\begin{center}Abstract\end{center}
Visual models play a crucial role in both science and science communication. However, the distinction between mere analogies and mathematically sound graphical representations is not easy and can be misunderstood not only by laypeople but also within academic literature itself. Moreover, even when the graphical representation exactly corresponds to the mathematical model, its interpretation is often far from obvious. In this paper we discuss the potential landscape visualization commonly used for tipping points in the context of nonlinear dynamics and reveal potential pitfalls, in particular when distinguishing bifurcation induced tipping (B-tipping) from noise-induced tipping (N-tipping).

We propose new visualization techniques for tipping dynamics, carefully distinguishing between B- and N-tipping as well as between single systems and ensembles of systems. Explicitly, we apply these visualizations both to molecular cell biology and to climate science in order to reveal the crucial differences in the interpretation of the visual models. We find that it is crucial to explicitly discuss the assumptions made within the visual model and to be aware of the risk of misinterpretation. These findings apply to a wide range of readership, from graduate students - as some general knowledge of nonlinear systems is required - to research professionals working in the field of nonlinear sciences. This paper provides the theoretical groundwork for these new visualizations. As a next step, we propose to investigate the individual mental models that might be induced by these visualizations using empirical research that builds upon these findings.\\


\section{Introduction}\label{intro}

In both science and science communication, visualizations play a crucial role in gaining intuition for a given physical process. Many publications, aimed at both professionals and laypeople, include visualizations to make the abstract and often complex physical systems more accessible and understandable \cite{Steffen2018, McKay2022, IPCC6PSB, Balazsi2011, Ferrell2012, Moris2016, Hao2018, Boeing2016, Dr2014}. From a didactic point of view, several aspects have to be taken into account when considering visualizations. First, a distinction must be made between the visualization presented by the author of a given publication with a specific \textit{purpose} and the individual \textit{mental model} the recipient might form based on this representation, which may be very different from the author's intended purpose. 

To analyze the potential pitfalls of visualizations, it is helpful to use a theoretical framework that distinguishes between different aspects of a viewer's interpretation. Pluta et al. \cite{Pluta} provide general criteria for "good models," which we can adapt to visualizations: they should have high levels of conceptual coherence and clarity, should be compatible with visualizations in other fields, should be appropriately parsimonious, and should be consistent with empirical evidence. 
It is important to note that the recipient's mental model can not only depend on the visualization, but also on their prior knowledge. The Students' Understanding of Models in Science instrument (SUMS) was developed by Treagust et al. \cite{Treagust} to gain insight into these mental models. Using exploratory factor analysis, Treagust et al. identified five factors, namely models as "multiple representations", as "exact replicas" and as "explanatory tools", the "uses of scientific model", and the "changing nature of models". Later, Ubben and Heusler \cite{Ubben2021} combined these instruments with context-specific questions related to models of the atomic shell.

Using exploratory factor analysis, two main factors were identified with regard to this context: 
The first factor is \textit{Fidelity of Gestalt (FG)}, which describes the extent to which models are understood to be accurate visual representations of phenomena or accurate representations of how things look, and the extent to which the gestalt of the models is perceived to be accurate. 
The second factor is \textit{Functional Fidelity (FF)}, which describes the extent to which models are understood as appropriate descriptions of how phenomena work, and the extent to which the underlying abstract functionality of the models is perceived as accurate.

This FF/FG distinction is crucial because many pitfalls in scientific visuals arise from a "category error," where a viewer interprets a functional model with high fidelity of gestalt. This framework, which has been empirically confirmed in various areas of physics \cite{Ubben2022, Ubben2023}, will guide our critique of existing visualizations and our proposal for new ones. It also provides the necessary vocabulary for the future empirical research we advocate for in our conclusion.

In this paper we focus our discussion on nonlinear dynamics for several reasons. Applications of nonlinear dynamics range from climate science to cell dynamics to models of brain activity and artificial intelligence, making it one of the most important frameworks in contemporary physics. For this reason, the communication of scientific results to the general public is particularly important in this context.
The inherent difficulty of conveying counter-intuitive concepts such as "bifurcations", "stochasticity", and "tipping points" with words alone makes visualizations an essential, yet challenging, communication technique. One of the most popular visualization techniques used to convey the key ideas of nonlinear dynamics is a "potential landscape" (see Fig.\ref{Bistable}). The "point particle" (which represents the state of the system) moves towards a "stable" minimum within the potential. While the strong point of this visualization is its universality, there is a price to pay for that: In each application at hand, the interpretation of the visualization is rather different. What is the meaning of the "potential"? How should we interpret "the point particle"? What mechanisms may change the behavior of the ball, in particular, the transition from one minimum of the potential to another (also called "tipping behavior")? 

Particularly in the context of climate science, the topic of "tipping points" attracts much attention \cite{IPCCAR6, Lenton2008}. Climate science seeks to understand the dynamics of the components of the Earth's climate system and to model and predict future changes in the Earth's climate \cite{Bader2008}. These climate components (e.g., atmosphere, AMOC, Greenland Ice, Monsoons, Amazon rainforest) are highly complex systems that also exhibit a wide range of nonlinear properties, such as stochastic and chaotic behavior, hysteresis and tipping points. Therefore, visualization techniques are ubiquitous to simplify the complex behavior of climate systems and to help understand and communicate findings effectively. For example, many major publications on climate tipping points also use potential landscape visualizations to explain system dynamics \cite{Scheffer2001, Lenton2008, Lenton2012, Steffen2018, IPCC6PSB, Lenton2023}. 
However, the same visualization ideas can also be applied to topics such as the transition of a cell from a pluripotent stem cell to different possible somatic cell states (called differentiation) in the field of molecular cell biology, where nonlinear dynamics also play an essential role \cite{Waddington2014, Balazsi2011}. Many publications in this field also rely on the use of potential landscape visualizations to explain and communicate these differentiation and cell fate decision processes \cite{Waddington2014, Wang2008, Balazsi2011, Ferrell2012, Moris2016, Brackston2018, Coomer2022}.

\begin{figure}[!htbp]
    \centering
    \includegraphics[width=\columnwidth]{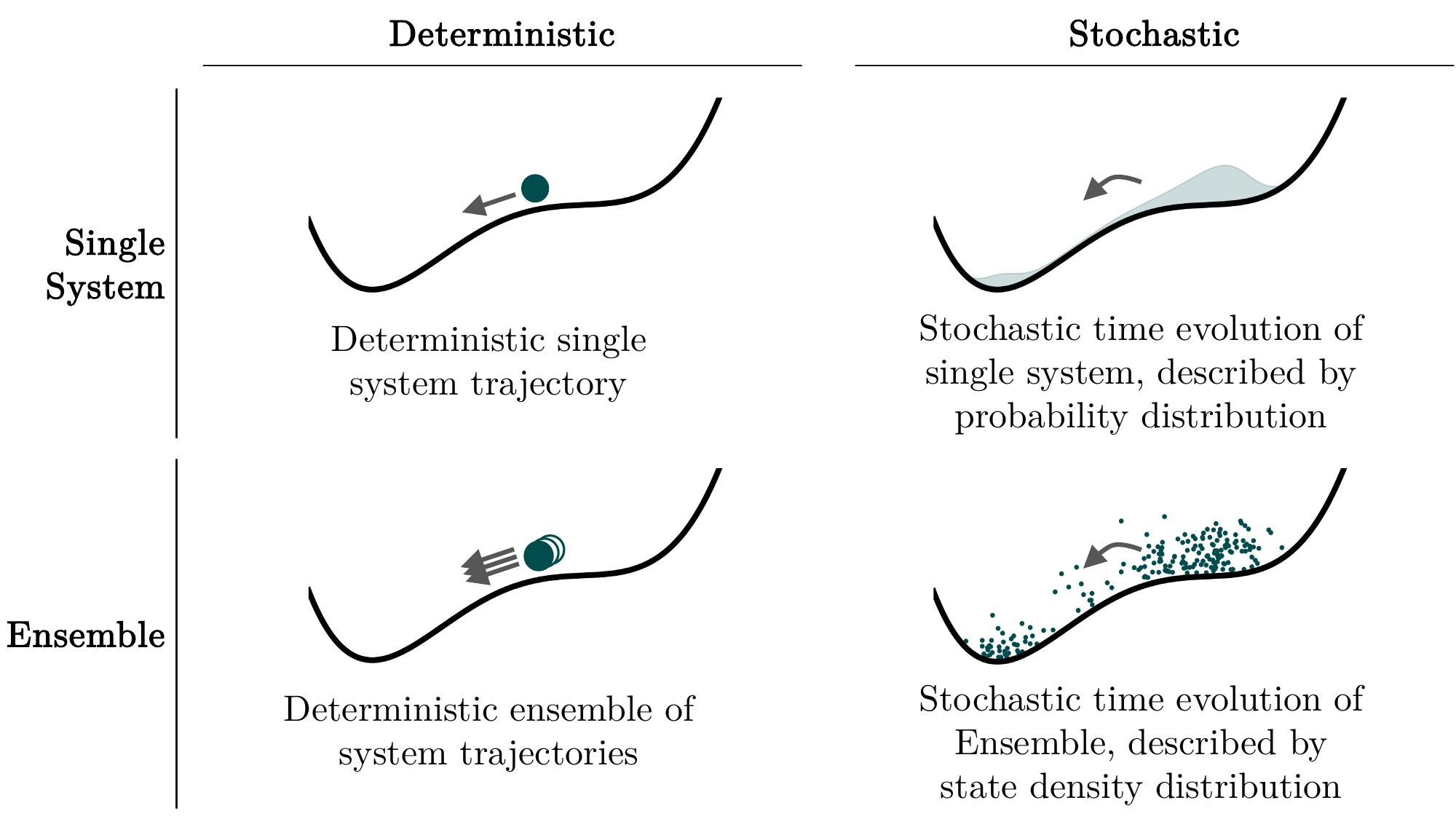}
    \caption{The interpretation of the visualization of the time evolution in the "potential landscape" differs, depending on whether a single system is considered or an ensemble, and whether the time evolution is deterministic or stochastic. The subtle differences in interpretation of these four different cases will be discussed in detail in this contribution.}
    \label{fig:FourField}
\end{figure}

We choose to focus on these two particular fields of science because on the one hand, both exhibit similar stochastic nonlinear behavior, in particular noise- and bifurcation-induced tipping. On the other hand, there are subtle differences in the interpretation of their respective  visualizations that we want to highlight (see Sec. \ref{sec:comparing}). As we will show later in this paper, visualizations of these systems can sometimes be ambiguous, misleading, and even problematic for various reasons. We will identify several misleading and problematic visualizations, either in their communicative purpose intended by the author, or in possible misinterpretations in terms of individual mental models, or even both. We then propose new visualization techniques for nonlinear stochastic processes from the fields of molecular cell biology and climate science as well as point out subtle difficulties in their interpretation. For that purpose, we carefully distinguish visualizations along two dimensions of interpretation: (i) interpretation as ensemble/individual system (ii) interpretation as deterministic/stochastic system (see Fig. \ref{fig:FourField}). While we apply our visualization to the context of climate science and cell dynamics, we argue that our visualization scheme is generalizable and may be adopted to any other context within the field of nonlinear dynamics. Further empirical research will be necessary to elucidate the mental models induced by these new visualizations.

This paper is organized as follows: First, we discuss analogies between classical mechanics and nonlinear dynamics, potential landscapes as a shared visualization technique and limitations of their analogies. We then give a brief introduction to the mathematical background of deterministic and stochastic nonlinear dynamics and in particular tipping behavior. More specifically, we distinguish between deterministic and stochastic systems as well as single and ensemble systems - and the resulting consequences for visualizations and their interpretation. Next, we focus on tipping behavior in climate systems and cell differentiation processes and  potential problems of their visualizations. We then propose our own visualizations, carefully distinguishing between deterministic and stochastic as well as single and ensemble systems. Finally, we suggest further empirical research to better understand the mental models these visualizations might induce, and emphasize the need for clear and accurate visual representations in scientific communication.

\section{The Analogy Between Classical Mechanics and Nonlinear Dynamics}
\label{sec:analogy}

\begin{figure}[!htbp]
    \centering
    \includegraphics[width=\columnwidth]{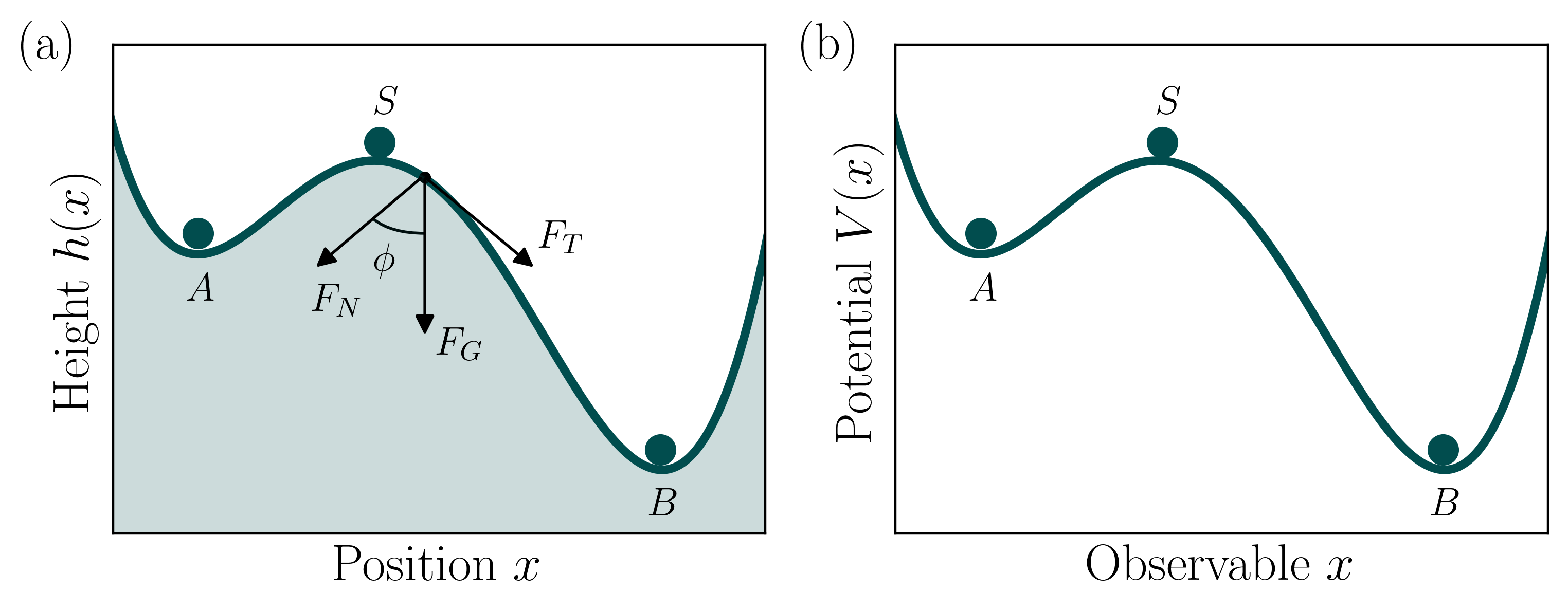}
    \caption{(a) Analogy from classical mechanics of a one-dimensional bistable system with valleys (stable equilibria) at points $A$ and $B$ separated by a saddle (instable equilibrium) at point $S$. (b) Abstract functional potential of the same shape with the same equilibria. Visualizations of this type are a common introductory metaphor for bistability and are found in many standard texts on nonlinear dynamics (e.g., Strogatz \cite{Strogatz})}
    \label{Bistable}
\end{figure}

Much of the intuition about nonlinear dynamics is based on analogies to classical mechanics (see e.g., Strogatz \cite{Strogatz}). In this context, consider the visualizations shown in Fig. \ref{Bistable}, which can be found in similar form in popular and academic literature. 

The "potential landscape" shown here describes the (potential) energy of a system as a function of one variable with the position of the system state, represented as spheres within the landscape, corresponding to the current energy of the system. If the potential landscape is understood by the recipient not as an abstract functional object (\textit{functional fidelity}), but as a real shape of a mountain landscape (\textit{fidelity of gestalt}), misinterpretations can arise. For instance, while frictionless motion in an abstract \textit{quadratic potential energy well} $V(x) \propto x^2$ indeed results in simple harmonic oscillation for any amplitude, this is not true if one misinterprets the visualization as a real \textit{physical track with a quadratic shape} $h(x) = ax^2$. In the latter case, the motion is constrained to the curve and the restoring force along the path becomes a nonlinear function of position, which deviates from simple harmonic oscillation for large amplitudes.
The constraint in the equation of motion introduced by the functional relation $z=h(x)$, can be treated using Lagrange parameters. The force 
\bea
    F_T = m g \sin(\phi) =  m g \frac{h'}{\sqrt{1 + (h')^2}}\quad \text{with}\quad  h' = dh/dx = \tan(\phi)
\eea
would then be nonlinear. On the other hand, if we took "fidelity of gestalt" literally, we could determine the shape $h(x)$ of a real "mountain landscape" that would indeed lead to a harmonic oscillation. For this we must require that the force $m g \sin(\phi)$ is proportional to the path length $s = \int \sqrt{dx^2+dz^2}$. This results in the following functional equation for $z=h(x)$:
\bea
\frac{h'}{\sqrt{1 + (h')^2}} = D \int_0^s \sqrt{1 + (h')^2} dx
\eea
with some proportionality constant $D$. 
While this is an amusing and perhaps unexpected problem of classical mechanics, it is obvious that the naive (FG)-interpretation of the potential as some real "mountain landscape" is a dead end, as the visual shape of a landscape and its abstract physical function are not the same. Instead, this analogy should be understood only in an abstract, functional sense (FF).

Even on this level, there are various interpretations of the "potential": In Newtonian mechanics, the potential energy (with the physical unit "Joule") is related to the physical concept of "work" in a conservative force field $\bm{F}$ via $V(\bm{x}) =  - \int^{\bm{x}} {\bm{F}} d{\bm{s}}$, equivalently, 
\bea
{\bm{F}}  = - {\bm\nabla} V({\bm{x}}) = m \frac{d^2 {\bm{x}}}{dt^2}.
\eea
Again by analogy, the concept of "potential" is further generalized when we consider models in nonlinear dynamics. Now the "state" of the system is no longer a point particle like in classical mechanics, but can have a vast number of interpretations - for example as a particular state of cell dynamics, brain activity, strength of the AMOC, percentage of forest cover of the Amazon rainforest, etc. For deterministic processes of the type 
\bea
\label{eq:Gradient}
{\dot {\bm{x}}} = \bm{f}({\bm{x}}, t, \eta) = - \nabla V({\bm{x}}, t, \eta),
\eea 
the interpretation of a "particle" (a system) at position ${\bm{x}}$ (a specific system state) in a "potential" $V({\bm{x}}, t, \eta)$ describing the dynamics of the system state for a given set $\eta$ of external parameters can be visualized as shown in Fig \ref{Bistable} (b). At any given time, the state is described by the vector ${\bm{x}}$ describing the unique and deterministic system state driven by the dynamics. Fixed points of the dynamics, where $\dot{\bm{x}}= f({\bm{x}}, t, \eta) = 0$, are represented by extremal points of the potential, with stable fixed points corresponding to minima. 
Remarkably, some general mechanisms can be derived that underlie all these different types of dynamics, revealing a certain universality of nonlinear dynamics. From this point of view, it is clear that modeling in terms of "potentials" must be seen as a rather abstract concept.

\section{Mathematical Background of Nonlinear Dynamics and Tipping Behavior}
    \label{sec:tipping}

    A tipping point can generally be defined as a threshold of a system where a small perturbation of the system state is sufficient to tip the system from one fixed point of the dynamics (minimum of the potential) to another fixed point \cite{Gladwell00, Lenton2008, Scheffer2009book}. This transition is usually caused by external forcing with a self-sustaining positive feedback loop that causes the tipping to continue even after the external forcing ceases. It is often abrupt and/or irreversible within a certain time frame, and may exhibit hysteresis behavior. Complex dissipative systems in the real world typically have at least one stable state (or attractor) to which they will return upon small perturbations. Two of the main mechanisms by which tipping can occur are bifurcation-induced tipping (B-tipping) and noise-induced tipping (N-tipping) \cite{Ashwin2012}.
    
    In bi- or multistable systems, i.e., systems that have more than one stable state, B-tipping can occur when one of the stable states disappears, changes its stability, or the system is forced into a qualitatively different stable state due to a local bifurcation. Bifurcations are caused by internal or external parameter changes corresponding to a change of the physical conditions of the system or its environment. In terms of the potential landscape, let $\bm{x}$ be the system state and $\eta$ an external forcing parameter. The values $(\bm{x_b}, \eta_b)$ at which the bifurcation occurs are called bifurcation points. Within the potential landscape, a stable minimum can then become unstable once this bifurcation points is reached and surpassed. 
    
    In contrast, N-tipping can occur when the current attractor of a system decreases in stability. The system is then more susceptible to small perturbations caused by stochastic fluctuations, i.e., noise. These can cause the system to leave the neighborhood of its stable attractor and cross the stability threshold, resulting in a transition to another stable state, even if the potential itself does not change its shape. N-tipping is caused by stochastic forcing, which drives the system state from the neighborhood of a stable fixed point across a stability boundary into the basin of attraction of another stable fixed point. To model the stochastic forcing of a dynamical system, an additive, state-independent noise term can be included in the model, resulting in the stochastic differential equation (SDE) \cite{Horsthemke2006}
    \bea
        d\bm{x} = \bm{f}(\bm{x}, \eta, t)dt + \bm{\sigma} d\bm{W}_t.
    \eea
    Here, the added term describes the stochastic forcing, where $\bm{\sigma}$ is the noise amplitude matrix and $d\bm{W}_t$ is the multidimensional increment of a Wiener process, i.e., white noise. Since the system is now non-deterministic, individual system trajectories are no longer sufficient. Instead, we must consider the evolution of the probability density function  $p(\bm{x}, \eta, t)$. The mathematical tool that governs the dynamics of this probability density is the Fokker-Planck equation \cite{Horsthemke2006}. While the full formalism is complex, we present it here briefly to mathematically ground our subsequent discussion of stochastic visualizations. The Fokker-Planck equation with the diffusion Matrix $\bm{D}=\frac{1}{2}\bm{\sigma\sigma}^{\text{T}}$ is 
    
    \begin{equation}
        \label{eq:fpe}
        \frac {\partial p(\bm{x} ,\eta,t)}{\partial t}=
        -\sum _{i=1}^{N}{\frac {\partial }{\partial x_{i}}}\bigl[f _{i}(\bm{x} ,\eta, t)\,p(\bm{x} ,\eta,t)\bigr]+
        \sum _{i=1}^{N}\sum _{j=1}^{N}{\frac {\partial ^{2}}{\partial x_{i}\,\partial x_{j}}}\bigl[D_{ij}\,p(\bm{x},\eta,t)\bigr].
    \end{equation}
    One of the crucial questions we want to address in this paper is: How can stochastic behavior be incorporated into the visualization of the potential landscape? As a preparation to this question, we again recall the original analogy to classical mechanics.  In many-particle physics and in thermodynamics, even if the underlying classical dynamics is deterministic, it is natural to introduce statistical methods to evaluate mean values and variations. The concept of a phase space density $\rho({\bm x}, {\bm p}, t)$ is central to this statistical description of time dynamics. Adding noise to a single-particle system can effectively describe fluctuations due to random interactions in many-particle systems, and we can therefore reduce its many-particle interaction to a model of a single particle with noise. By analogy, we can also add noise to nonlinear systems as an effective description of unknown interactions with the environment. Now the state of the system can only be described by a probability density $p({\bm{x}}, t)$, like in a diffusive process. For this reason, the interpretation of the visualization of the system state as a single "point particle" (as shown in Fig. \ref{Bistable}) becomes undefined for N-tipping, and indeed this visualization is imprecise since it does not show the stochastic nature of the process.
    
    Dynamical systems in the real world are almost always subject to internal or external forcing, leading to possible interactions between B- and N-tipping. Small fluctuations of the system state will not lead to tipping as long as the shape of the potential landscape does not change and the potential well is much higher than the magnitude of the fluctuation. However, when the system approaches a bifurcation point caused by a change in the external parameters $\eta$, the system loses resilience (i.e., the potential well flattens) and small fluctuations can then cause tipping even before the bifurcation point is reached. In the case of strong forcing and large fluctuations, tipping becomes possible even far away from a bifurcation point.
    
    To further illustrate this, we consider the simple system of a single particle under stochastic forcing (also known as "Brownian motion"). In classical statistical mechanics, this system is interpreted as a point particle, and in nonlinear dynamics as a system state. Solving the diffusion equation for an initial state of the "particle" at $x=0$ yields the well-known bell-shaped probability density $\rho(x,t)dx$ for detecting it in the interval $[x, x + dx]$ that diffuses with time. For a single "particle", its exact state cannot be determined, only the probability distribution of the particle (or system state) is known. However, in an ensemble interpretation of a large number $N$ of particles (or systems) we can calculate expectation values with $N \rho(x, t) d x$ and for large $N$ the fluctuations will decrease with $1/\sqrt{N}$. It is \textit{only} in this limit of a large number of individual systems that the ensemble interpretation becomes meaningful and definitive statements about expectation values can be made. Therefore, a careful distinction between the single particle interpretation and the ensemble interpretation is \textit{crucial}.
    
\begin{figure}[!htbp]
        \centering
        \includegraphics[width=\columnwidth]{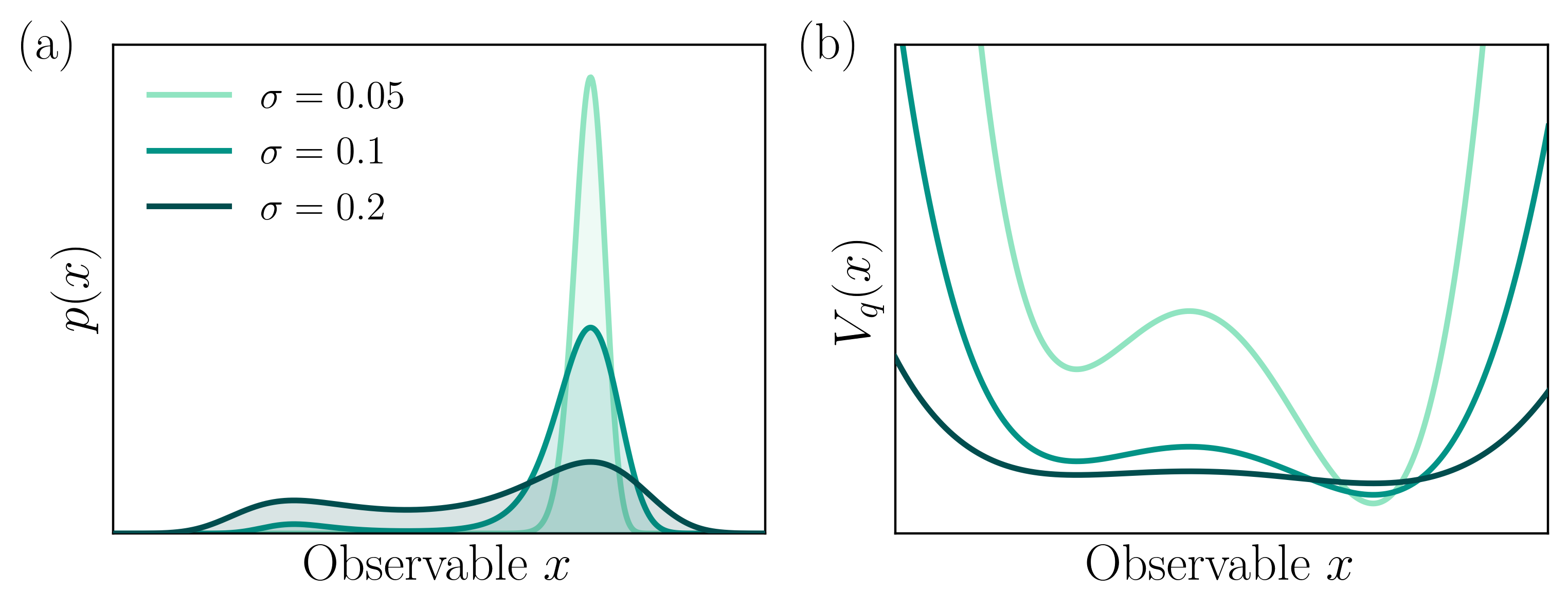}
        \caption{(a) Probability density function (PDF) $p(x)$ of a one-dimensional system with an example potential under additive stochastic forcing with varying noise amplitudes $\sigma$. (b) Quasi-potentials $V_q(x)$ calculated from the PDF for the system under the same stochastic forcing.}
        \label{Qpot}
    \end{figure}

    It can be challenging to incorporate stochastic forcing into the potential picture. In literature, one possible visualization technique is the introduction of a "quasi-potential" $V_q(\bm{x}, \eta, t)$ \cite{Zhou2012}. The depth of a well of the quasi-potential is related to the probability density function $p(\bm{x}, \eta, t)$ of the system by 
    \bea
         V_q(\bm{x},\eta, t)\sim-\ln p(\bm{x}, \eta, t),
    \eea
    reminiscent of the Boltzmann distribution from statistical physics.
    In the one-dimensional case with an state-independent additive noise term, the Fokker-Planck-Equation simplifies to 
    \bea
         \frac{\partial p(x,\eta,t)}{\partial t} =-\frac{\partial}{\partial x}\bigl[ f(x,\eta, t)\, p(x,\eta, t)\bigr]+\frac{\sigma^2}{2}\frac{\partial^2 }{\partial x^2}p(x,\eta, t).
    \eea
    Rewriting its solution to include the "standard" deterministic potential $V(x,\eta, t)$, yields the probability distribution
    \bea
        p(x,\eta, t) = \frac{C}{\sigma^2}\exp{\left(-\frac{2V(x,\eta, t)}{\sigma^2}\right)},
    \eea
    and thus the quasi-potential 
    \bea
        V_q(x,\eta, t) = \frac{2}{\sigma^2}V(x,\eta, t) + \tilde{C},
    \eea
    with the constants $C, \tilde{C}$. Here, an increase in noise amplitude $\sigma$ effectively leads to a flattening of the quasi-potential, symbolizing the higher probability of the system state crossing a potential threshold as can be seen in Fig. \ref{Qpot}.
    
    However, the quasi-potential $V_q$ is conceptually very distinct from the deterministic potential $V$. The deterministic potential exactly determines the rate of change and thus the deterministic dynamics of a system. This is true for a single system as well as for an ensemble of systems. The stochastic quasi-potential, on the other hand, does \textit{not} allow any definitive conclusions about the trajectory of a single system, only about its probability distribution. Here, the single and ensemble case need to be distinguished carefully. Only in the limit of a large number of ensemble members the interpretation as a state density of the ensemble states becomes a meaningful interpretation. This is an important example of the distinction between the intended purpose of a visualization and the mental models it may induce: Using the same visual language ("potential landscape") for completely different scenarios obviously carries the risk of misunderstanding, even for those recipients who are aware that fidelity of gestalt (FG) must be abandoned in order to grasp the meaning of the visualization shown in Fig. \ref{Qpot} (b). It is for this reason that we explore alternative representations, which is a central contribution of the present work. 
    
    When discussing visual representations of tipping behavior, in the context of B-tipping, refined representations of bifurcation diagrams should also be discussed in a similar manner. Theoretically, the same distinction between ensemble and single system must be made as for the potential landscape, and the same difficulties exist in representing stochastic influences. However, we argue that bifurcation diagrams are significantly less prone to misinterpretation for a number of reasons. First, there is no real-world equivalent to a bifurcation diagram, as there is with the potential landscape, which reduces the likelihood of incorrect (FG) interpretations. For this reason, the diagram cannot be intuitively interpreted as analogous to a real physical object (see section \ref{sec:analogy}), but only as an abstract functional object, which makes correct interpretation more likely. Second, bifurcation diagrams are more often used in publications aimed at domain experts, who are less likely to misinterpret them. Visualizations of nonlinear dynamics using potential landscapes are often used in literature aimed at non-experts or laypeople, which makes a discussion of possible misunderstandings even more urgent. For these reasons, we will focus solely on potential landscape representations in this paper.

\section{Visual Models for Stochastic Nonlinear Processes: Comparing the Single System Interpretation with an Ensemble Interpretation}
\label{sec:comparing}

    In this section, we propose our own visualization technique, carefully distinguishing ensemble from single systems and stochastic from deterministic behavior, as shown in Fig. \ref{fig:FourField}. Later on, we apply these visualizations to cell dynamics and to tipping points in the climate system. First, we extend the simple example of Brownian motion to a more complex nonlinear system with tipping behavior. Here, the same distinction between individual systems and ensembles as well as their interpretations must be made.
    
\begin{figure}[H]
        \centering
        \includegraphics[width=0.5\columnwidth]{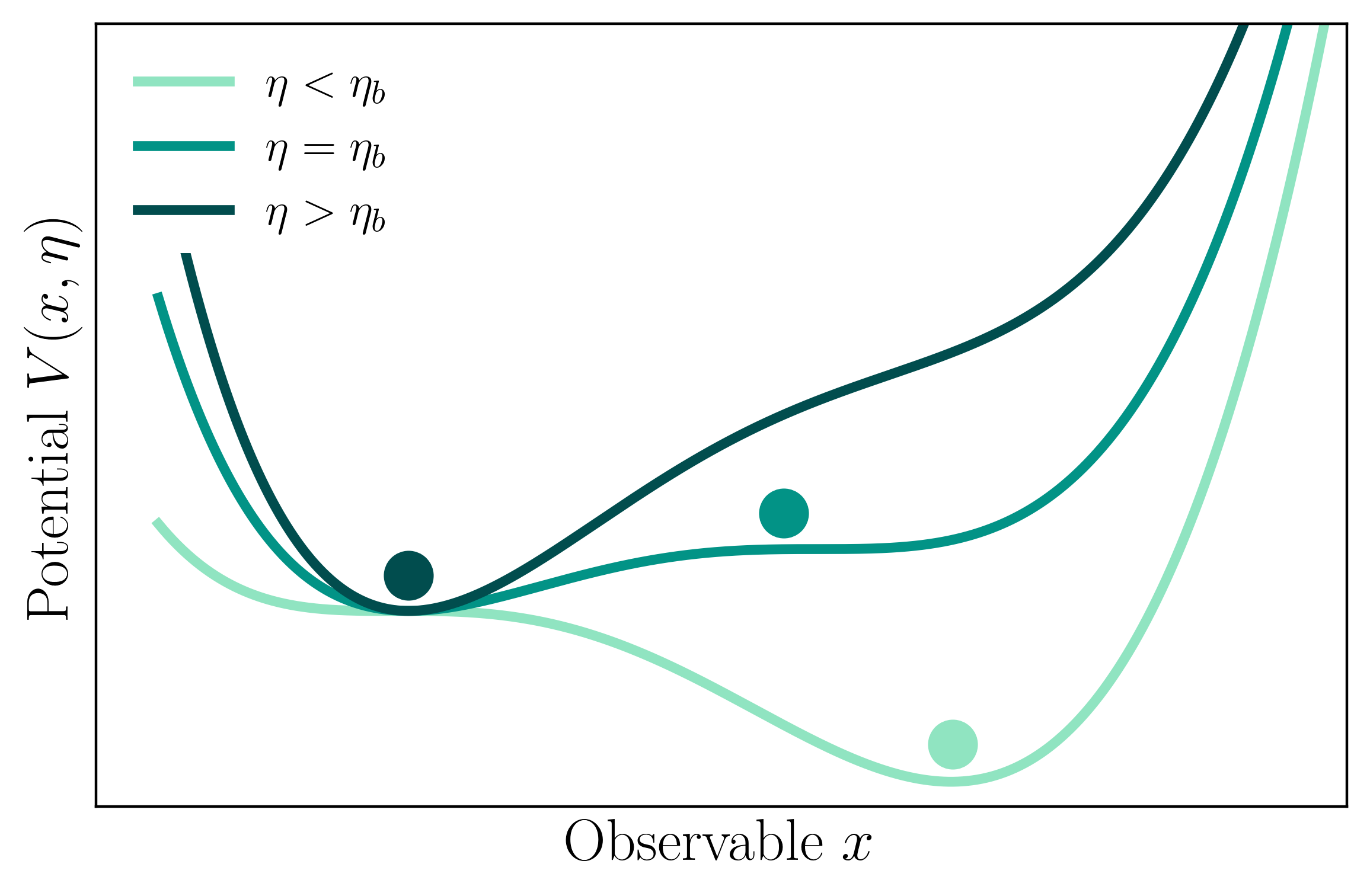}
        \caption{B-tipping of a deterministic system without stochastic forcing. The potential wells flatten out as the system approaches the bifurcation point from below ($\eta \rightarrow \eta_b$) and the system state transitions to a tipped state (to the left) after the tipping point is crossed.}
        \label{fig:deterministic}
    \end{figure}

    In the deterministic case without stochastic forcing, the ensemble behaves identically to a single system state, with each member trivially following exactly the same path determined by the underlying deterministic system dynamics. These cases are often visualized by a potential landscape in which the system state is represented by a sphere. B-tipping then leads to a change in stability, e.g., the flattening of a potential well, as visualized in Figure \ref{fig:deterministic}.
    \enlargethispage{3\baselineskip}

\begin{figure}[!htbp]
        \centering
        \begin{subfigure}[b]{0.49\columnwidth}
            \centering
            \includegraphics[width=\columnwidth]{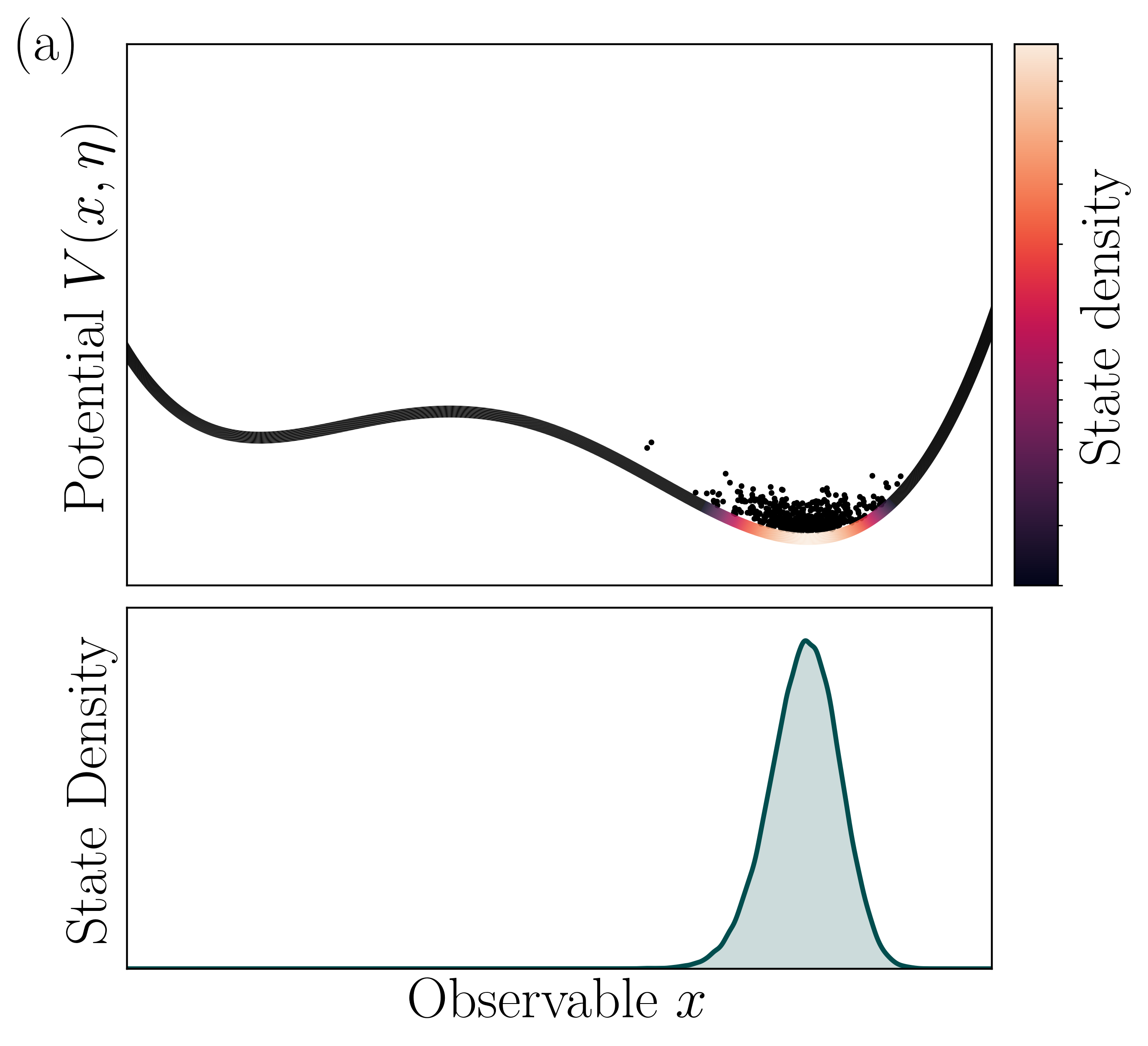}
        \end{subfigure}
        \begin{subfigure}[b]{0.49\columnwidth}
            \centering
            \includegraphics[width=\columnwidth]{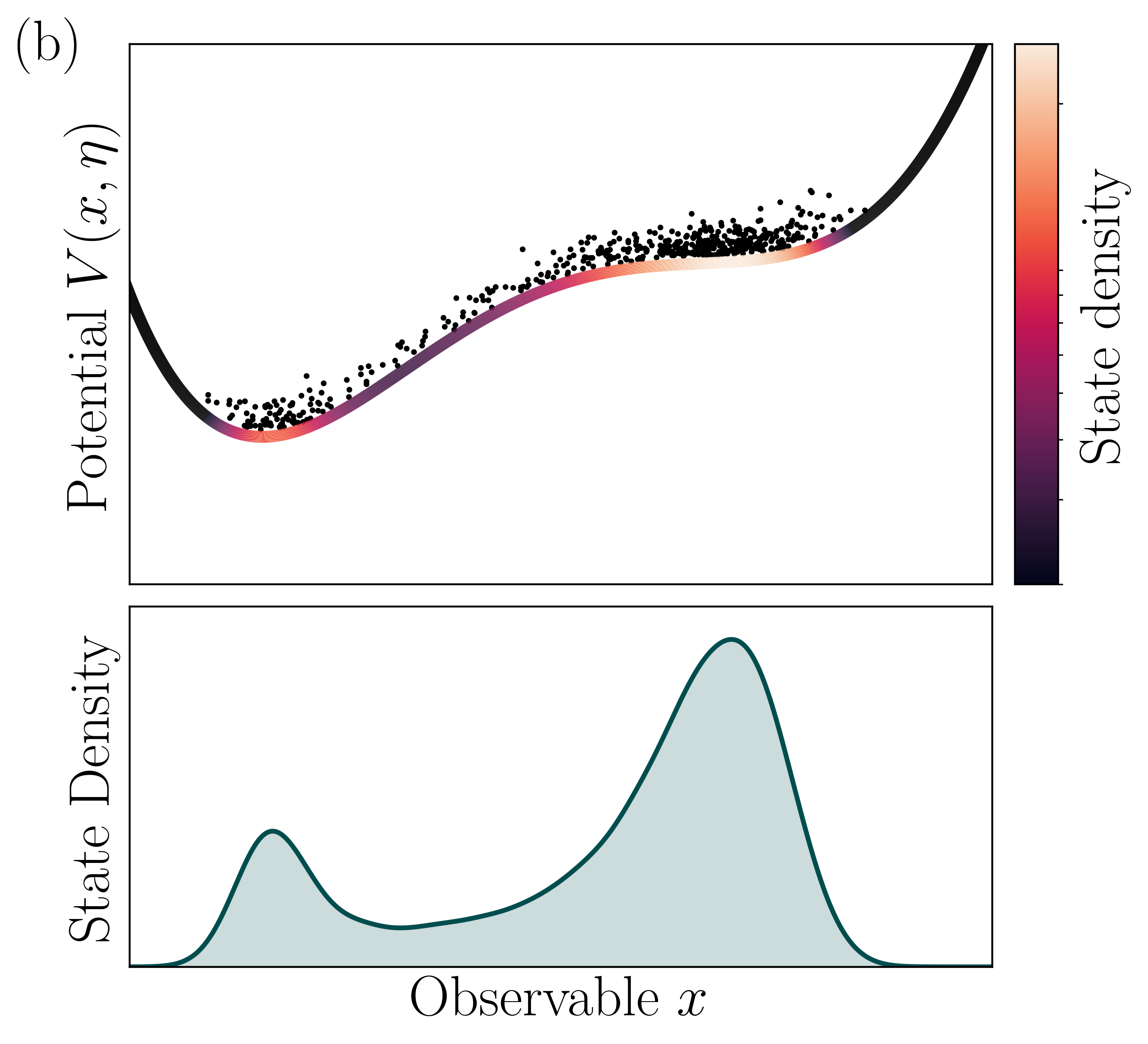}
        \end{subfigure}
        \par\bigskip
        \begin{subfigure}[b]{0.49\columnwidth}
            \centering
            \includegraphics[width=\columnwidth]{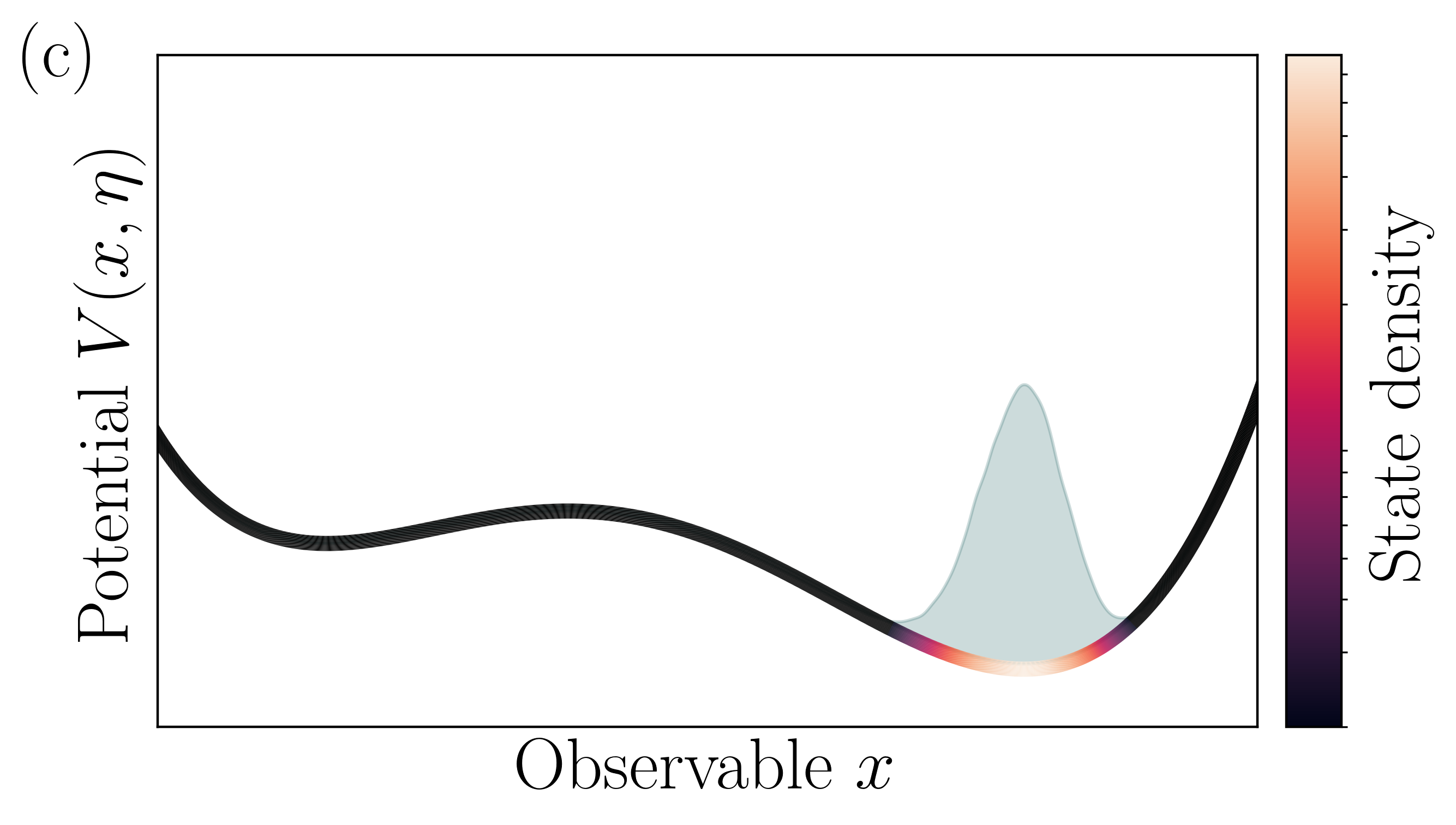}
        \end{subfigure}
        \begin{subfigure}[b]{0.49\columnwidth}
            \centering
            \includegraphics[width=\columnwidth]{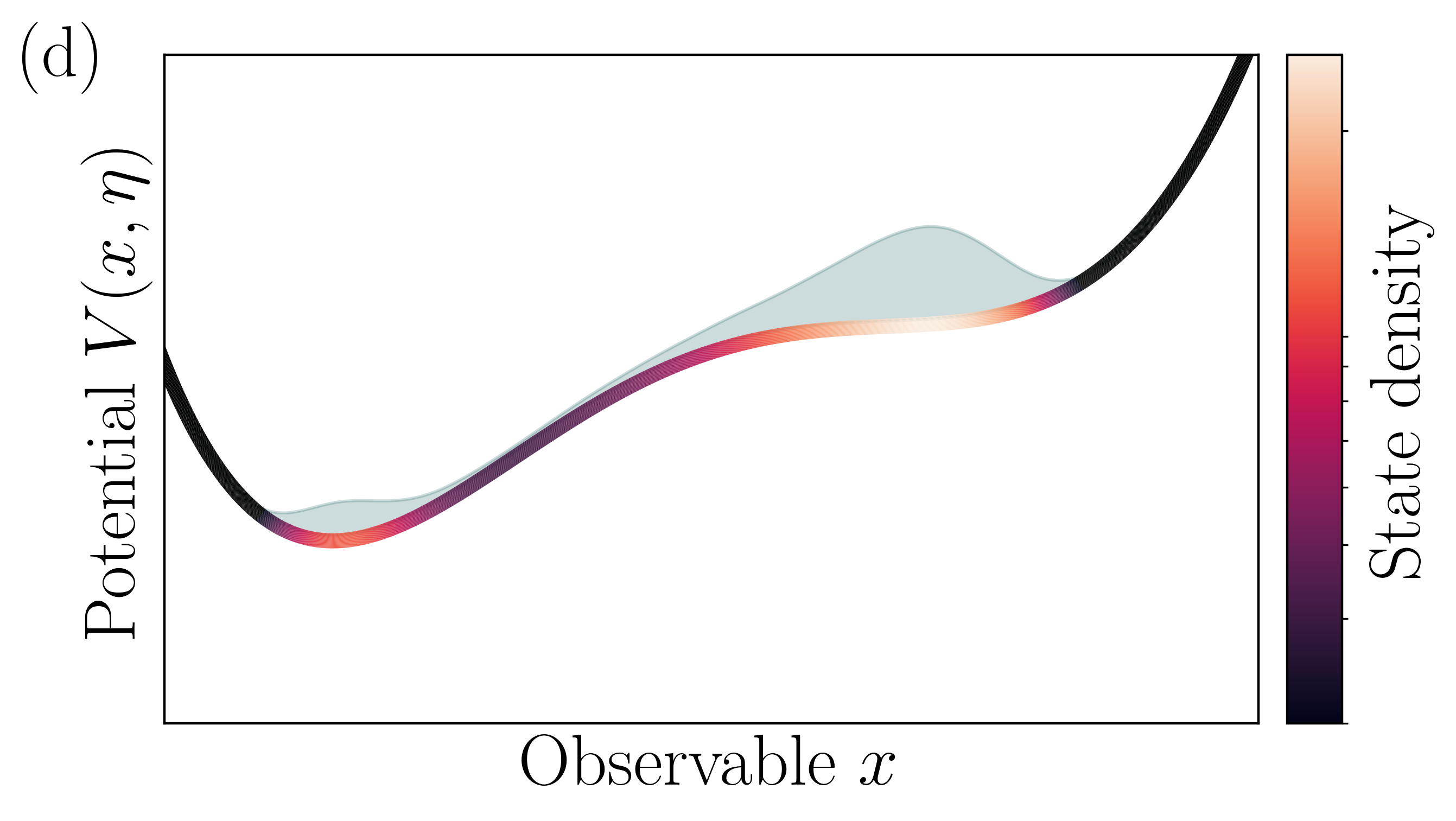}
        \end{subfigure}
        \caption{(a, b): Ensemble interpretation of a stochastic nonlinear process. (c, d): Probabilistic interpretation of the same dynamics. (a, c): Situation far from the bifurcation. (b, d): Situation near the bifurcation. Due to the stochastic forcing, there is a certain probability of N-tipping even before the B-tipping point is reached.}
        \label{Inter}
    \end{figure}

    With the introduction of stochastic forcing, an additional diffusion of system states (for an ensemble) or the probability density (for a single system) emerges. In Fig. \ref{Inter}, we present a potential landscape visualization of a nonlinear dynamical process under stochastic forcing (B- and N-tipping) for an ensemble of systems (a, b) and a single system (c, d). 

    It is important to note that these proposed visualizations are designed specifically to address the challenge of representing stochasticity and distinguishing between single-system and ensemble interpretations. While the underlying "potential landscape" metaphor is retained to provide a familiar visual scaffold, the key innovation lies in the representation of the system state itself. By visually differentiating between the cloud of points representing an ensemble's distribution of states (Fig. \ref{Inter}a, b) and the probability density shading representing a single stochastic system (Fig. \ref{Inter}c, d), our approach makes these crucial, and often overlooked, distinctions explicit. This is a critical improvement, as traditional visualizations often use a single, ambiguous sphere to represent the system state, regardless of whether it is a deterministic or stochastic system, a single system or an entire ensemble. Differentiating these cases is essential for accurately depicting the interplay between bifurcation-induced and noise-induced tipping.
    
    In an \textit{ensemble} under stochastic forcing, different individual system states diffuse to different fixed points of the dynamics, as shown in Fig. \ref{Inter}a, b. In this interpretation, each member of the ensemble can be said to behave differently, and we can characterize this behavior in the limit of large $N$ by a cloud of "particles", each representing an individual system. Fig. \ref{Inter} also shows the state density of the ensemble (which coincides with the probability density of the individual system in the limit of large ensemble size $N$). In this paper, as a prototypical example of the applicability of the ensemble interpretation, we discuss molecular cell dynamics, where each "particle" represents the state of a single cell.
    
    In contrast, Fig. \ref{Inter}c, d show the diffusion of the probability density within the potential landscape for a $single$ system. Here, since $N=1$, the ensemble interpretation is no longer possible. Instead, the probability density function that describes the state probability of a single system must be used. Our representation shows the probability distribution within the potential landscape. In this paper, as a prototypical example of the applicability of the stochastic interpretation of a single classical system, we consider a model of one component of the climate system, the Amazon rainforest.
    
\section{Tipping Dynamics and Popular Visualizations: From Cell Dynamics to Climate System Tipping Points}
    
    In the following, we argue that both the ensemble interpretation and the single system (i.e., probability) interpretation are important in real systems and must be carefully distinguished depending on the application. We discuss examples from literature regarding this distinction between deterministic/nondeterministic processes on the one hand, and the ensemble/probability interpretation on the other hand, and show that these subtle distinctions are not carefully made in all cases. 

    \subsection{Tipping points in cell dynamics}
\begin{figure}[htb]
            \centering
            \includegraphics[width=0.5\columnwidth]{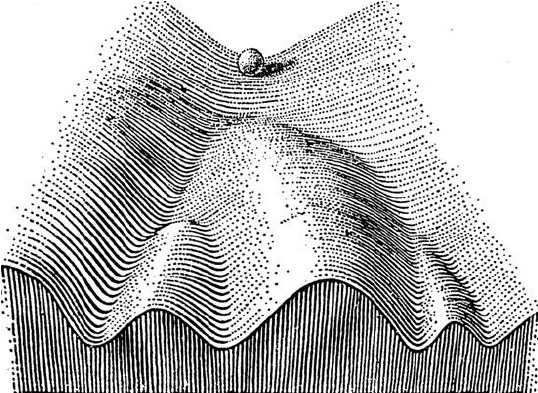}
            \caption{Waddington's original epigenetic landscape, adapted from Waddington \cite{Waddington2014}. }
            \label{WaddingtonLandscape}
        \end{figure}

        In 1957, the developmental biologist Waddington bridged the gap between theoretical and analytical biology by introducing the concept of an epigenetic landscape \cite{Waddington2014}. This visualization technique, derived from the theory of nonlinear dynamics, describes how a stem cell transitions from an undifferentiated state to one of its distinct cell fates during its development, illustrated by a sphere rolling through a landscape of diverging valleys (as shown in Fig. \ref{WaddingtonLandscape}). The shape of the landscape is influenced by the underlying complex regulatory networks.
        The point particle, representing the state of the cell, starts off in a single valley, the initial pluripotent stem cell. During development, it then rolls down the slope of potential, entering one of the valleys at each intersection, influenced by external stimuli or internal processes. It eventually ends up in one of the many sub-valleys, each representing a different somatic cell state. The $x$-axis represents the phenotype (simplified to a single state variable) and the $z$-axis (towards the viewer) can represent either time or a single input variable, its value increasing towards the viewer.
        In dynamic systems terminology, this concept can be expressed as a system state within a potential landscape subject to multiple occurrences of B-tipping. The cell fate decision points correspond to supercritical pitchfork bifurcation points, and the valleys correspond to fixed points in the phase space of the cell phenotype.
        
        Although widely used as a popular visual metaphor, Waddington's landscape has several shortcomings. While recent biological advances, such as transdifferentiation between different somatic states (i.e., direct transition between terminal valleys) challenge the hierarchy of cell states proposed in the epigenetic landscape, it also has shortcomings from a dynamical systems perspective. The landscape itself does not explain the emergence of variations in the fates of different cells, since without stochastic influences (N-tipping) all cells should follow the same deterministic path given the same initial condition. 
        
        Waddington's landscape metaphor has since been extended to include stochastic influences \cite{Coomer2022} and different types of bifurcations \cite{Ferrell2012}. An ensemble of identical undifferentiated stem cells undergoes the process of differentiation, while internal and external stochastic influences, such as probabilistic chemical reactions and fluctuations in their physical and chemical environment \cite{Coomer2022}, lead to a divergence of system states, resulting in cells with different phenotypes. From a dynamical systems point of view, an ensemble of systems with the same initial conditions subjected to multiple stages of combined B- and N-tipping will result in drift and diffusion of the individual ensemble members. It is important to note, however, that physically meaningful statements about the exact system states (position in the epigenetic landscape) of a single cell generally cannot be made, since its external and internal influences and interactions are described by stochastic forcing. Only in the limit of a large number of ensemble members the ensemble interpretation and thus the statistical treatment becomes meaningful.

    \subsection{Tipping points in climate systems}
        In the field of climate science, the possibility of "abrupt climate change" and "critical points" was considered as early as the 1980s \cite{CrowNorth1988}. The notion that anthropogenic influence could also trigger sudden transitions in climate systems, slowly gained traction in the early 2000s \cite{Lock2001, Sch1}.  One of the first comprehensive lists of potential tipping points in climate systems was compiled in 2008 by Lenton et al. \cite{Lenton2008}, and other climate components that may exhibit tipping behavior continue to be identified  \cite{McKay2022, Lenton2023}.
        
\begin{figure}[!htbp]
            \centering
            \includegraphics[width=0.5\columnwidth]{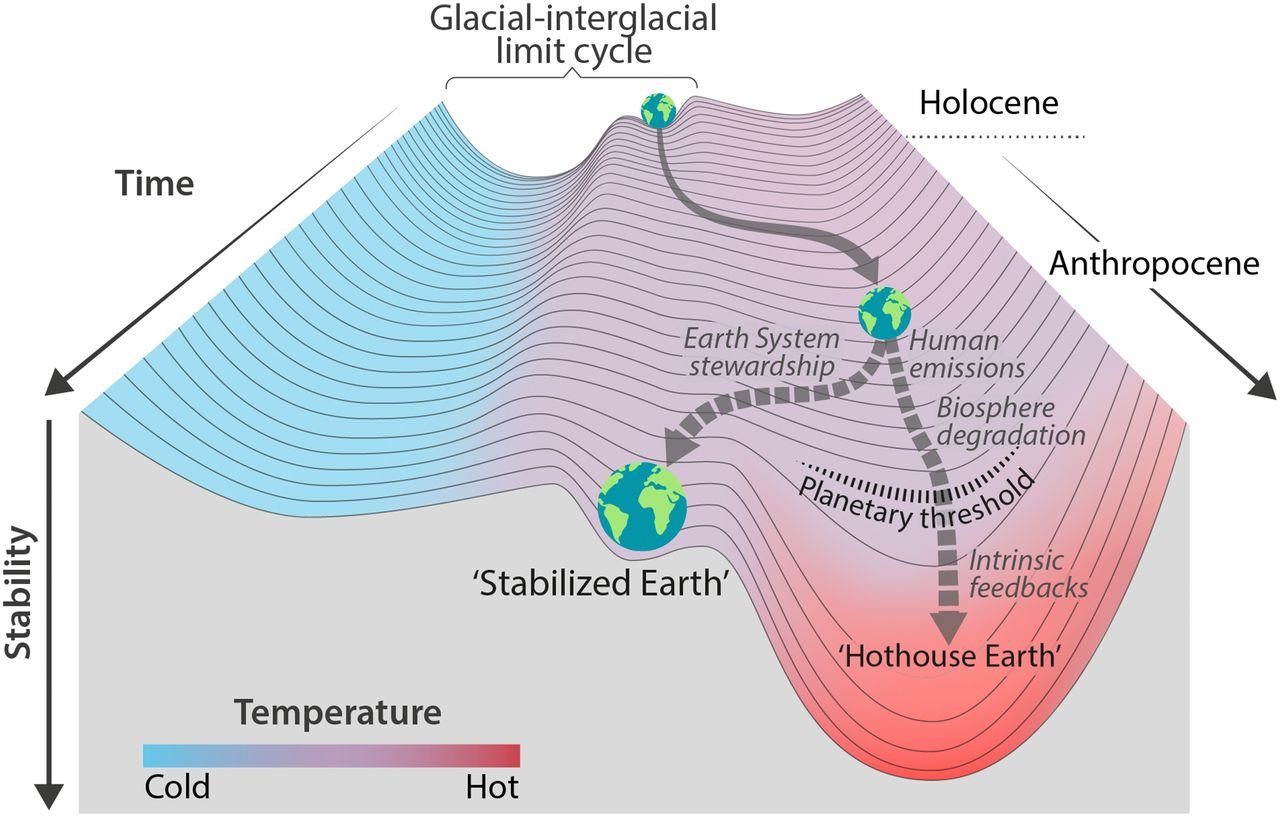}
            \caption{Visualization of different Earth trajectories under different global warming pathways, adapted from Steffen et al. \cite{Steffen2018}. Time increases along the $z$-axis towards the viewer.}
            \label{Lit2.2}
        \end{figure}

        Climate systems, such as the Amazon rainforest, the Greenland and Antarctic ice sheets, various monsoons or boreal permafrost, are complex dynamical systems that usually reside in a stable equilibrium determined by internal processes and boundary conditions. Changing external conditions, such as an increase in greenhouse gas emissions or rising global surface temperatures, can lead to qualitative changes in the stability landscape of the system, resulting in B-tipping. Tipping of climate systems can also be caused by stochastic forcing of the system state (N-Tipping), such as weather patterns and extreme weather events, and by human intervention in ecosystems. These are expected to have an even greater destabilizing effect on systems as global temperatures rise and the stability and resilience of climate systems decreases (see Interaction of B- and N-tipping in section \ref{sec:tipping}). Stochasticity can also be used to model the uncertainty introduced by the complexity of the physical processes involved in these climate systems (see section \ref{sec:tipping}). However, this uncertainty about the current and future state of the system can be difficult to incorporate into visualizations. Although the purpose of visualizations of the type shown in Fig. \ref{Lit2.2}, taken from Steffen et al. 2018 \cite{Steffen2018}, is clear, several points could be misunderstood at the level of the individual mental model that might be induced by the visualizations. 
        
        First, since the three-dimensional potential landscape is fixed in time, the model describes only \textit{one} particular greenhouse gas emission scenario in the future. The "decision point", where the trajectories of the Earth (obviously a single system rather than an ensemble) follow either the "Stabilized Earth" or the "Hothouse Earth" path, is labeled "Human Emissions". However, these emissions should change the shape of the potential landscape itself. Within a given emissions scenario, the trajectory is deterministic (without noise) or can only be described by a probability density. In summary, these visualization types do not clearly distinguish B-tipping from N-tipping, and do not clearly indicate that only a single B-tipping scenario is being shown, which in this case is equivalent to only a \textit{single} projection of future greenhouse gas emissions. 
        
        In the following, we propose alternative visualization techniques for these tipping processes in the aforementioned fields of molecular biology and climate systems using explicit mathematical models.

\section{Applications: From Cell Differentiation to Tipping of the Amazon Rainforest}
    \label{sec:visualization}
    
    To demonstrate how our proposed visualization framework can be applied to concrete physical scenarios, we now ground our discussion in two simple, established, one-variable models of system tipping. The first, from molecular biology, describes cell differentiation, an example of an \textit{ensemble} of systems. The second, from climate science, models the potential dieback of the Amazon rainforest, an example of a \textit{single} system. We present visualization methods incorporating uncertainty and stochasticity into the potential picture of the system, illustrating the practical importance of the conceptual distinctions made in this paper.

\subsection{Cell differentiation}
    To mathematically model the process of cell differentiation, we adapt a simplified model from Ferrell and James \cite{Ferrell2012} that models cell fate induction, the process by which cells commit to a particular lineage or function through signals from gene regulatory networks or neighboring cells. In our adapted model, the variable $x$ represents the synthesis of a differentiation regulator, the molecule responsible for promoting cell differentiation. These regulators are usually proteins such as transcription factors or signaling molecules that help "guide" the cell toward a particular fate and stimulate its own production. Their rate of synthesis is given by a baseline rate $\eta$ (the regulator is constantly produced at a certain rate), which serves as a bifurcation parameter. As more of the protein is present in the cell, even more of it is produced - a positive feedback loop which is modeled by a Hill function \cite{Gesztelyi2012}. The total synthesis rate is then given by 
    $\eta + \alpha \frac{x^5}{K^5+x^5}$, where the parameter $K$ represents the concentration of $x$ at which the response rate reaches half its maximum. $K$ essentially acts like a threshold - when $x$ approaches $K$, the feedback loop is strongly engaged. Additionally, the regulator molecule is constantly being broken down or removed from the cell at the so-called degradation rate $\beta x$, which is assumed to be linear.
    Combining these different terms results in the differential equation
    \bea
        \label{eq:ODE_cell}
        \frac{dx}{dt} = \eta + \alpha \frac{x^5}{K^5+x^5} - \beta x.
    \eea
    For simplicity, we follow Ferrell and James \cite{Ferrell2012} in assuming that the variables are dimensionless. With the parameter values $K=1$, $\alpha=1$, and $\beta=0.5$, the system exhibits a saddle-node bifurcation at $(x_b,\eta_b)\approx(0.58,0.23)$. 
    We assume that an inductive stimulus, such as an external signal like a hormone or interaction with other cells, leads to a slow increase in the baseline synthesis rate $\eta$, starting from a small initial value of $\eta=0.05$. When the synthesis rate of the regulator reaches and exceeds the threshold value $\eta \approx 0.23$, the undifferentiated state of the cell disappears at $x\approx0.58$, forcing the cell to adopt the differentiated state at $x\approx 2.44$. This means that the cell suddenly has to commit to a certain fate - a tipping process. 
    
    Combining an additive noise term to the equation \eqref{eq:ODE_cell} now allows us to model slight variations in the synthesis $x$ of the differentiation regulator caused by external influences or chemical/biological variations in the cell's environment. An ensemble of undifferentiated cells could then commit to a differentiated state at different stimulus intensities - or not at all.
    
\begin{figure}[H]
        \centering
        \includegraphics[width=0.5\columnwidth]{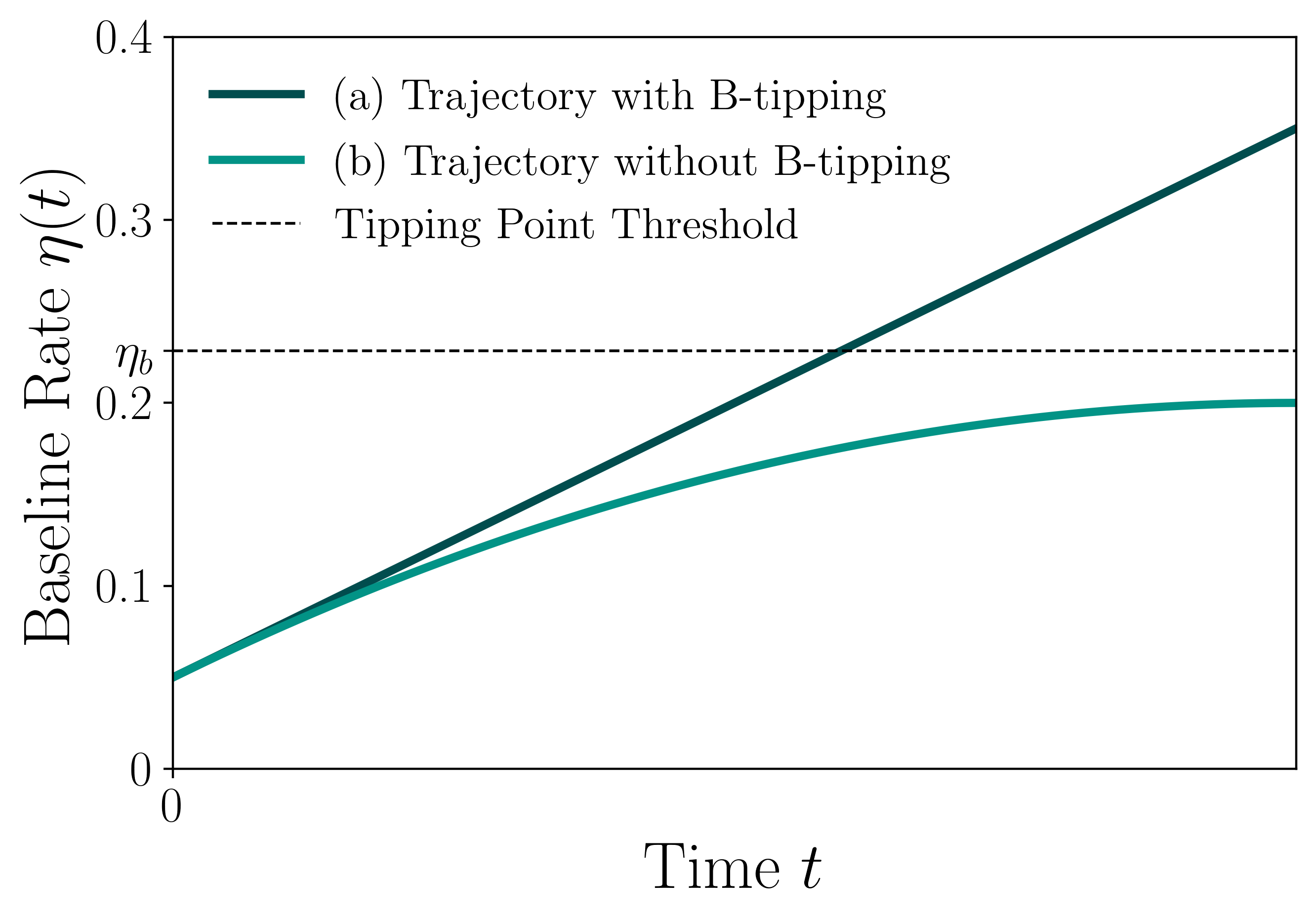}
        \caption{Bifurcation parameter $\eta(t)$ for the two different system trajectories visualized in Fig. \ref{Cells3D}. Trajectory (a) increases linearly crossing the B-tipping threshold while trajectory (b) approaches but does not cross it.}
        \label{fig:BifurcationParam1}
    \end{figure}

\begin{figure}[H]
        \centering
        \begin{subfigure}[b]{0.49\columnwidth}
            \centering
            \includegraphics[width=\columnwidth]{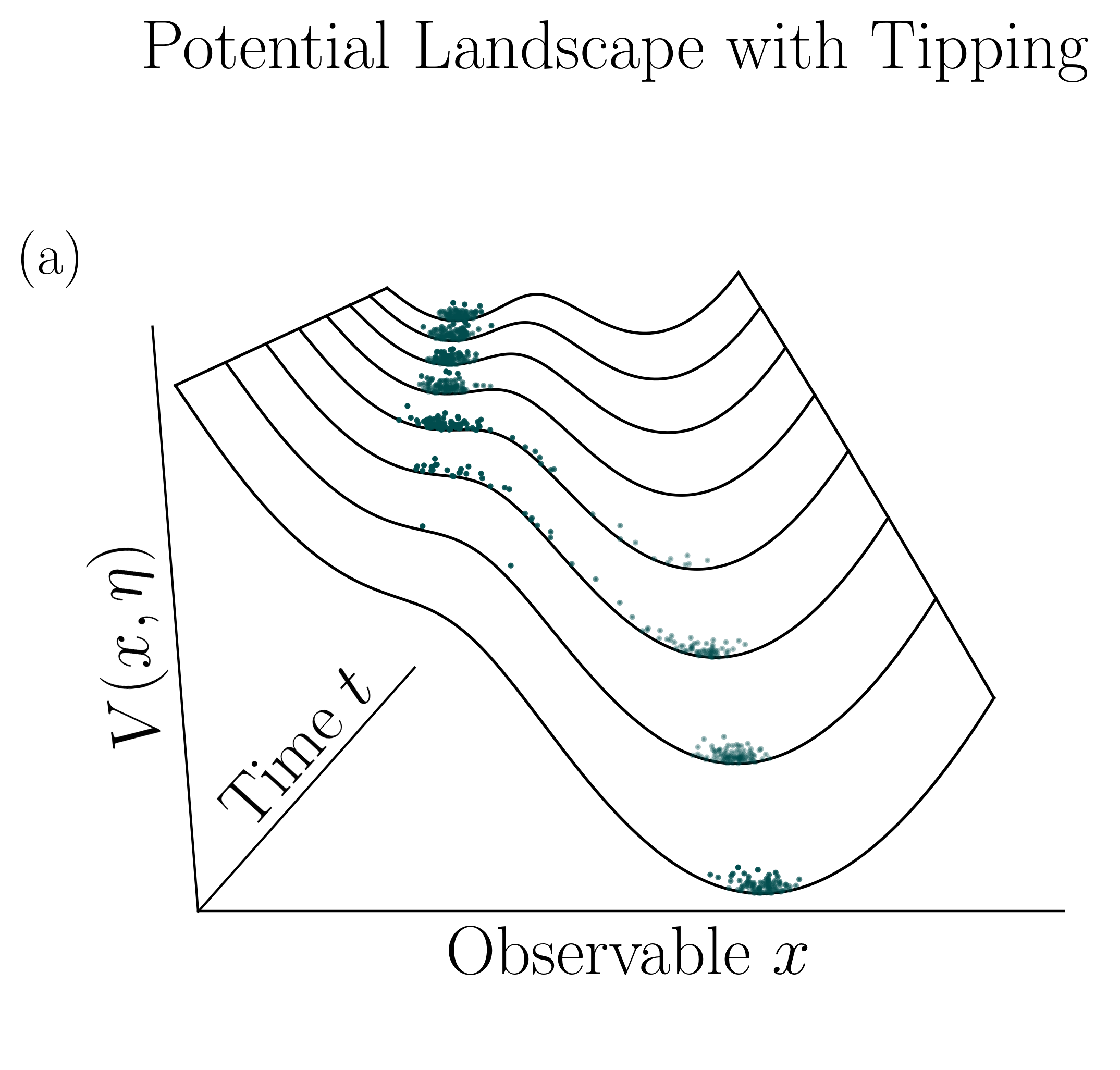}
        \end{subfigure}
        \begin{subfigure}[b]{0.5\columnwidth}
            \centering
            \includegraphics[width=\columnwidth]{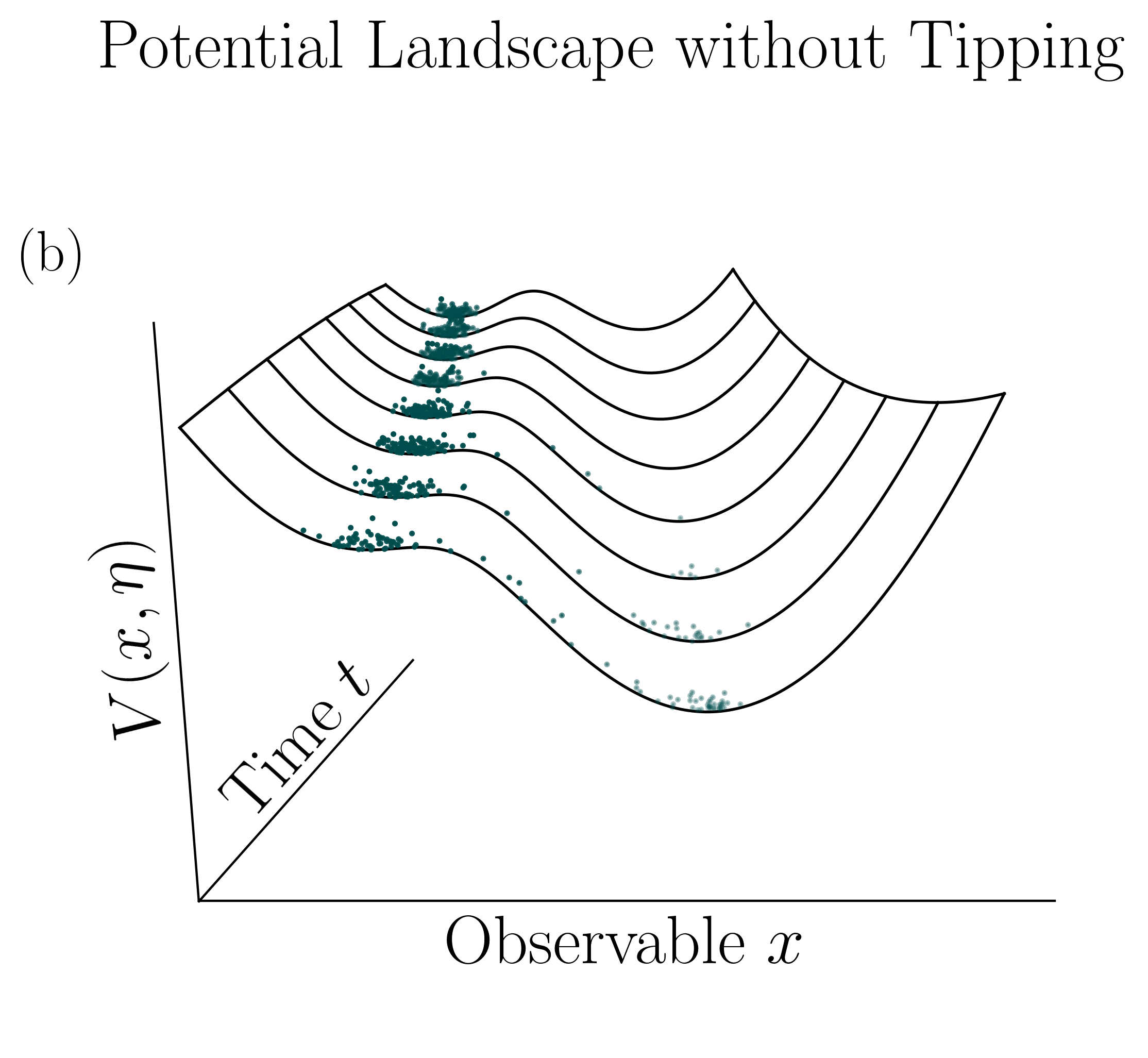}
        \end{subfigure}
        \caption{Potential visualization of two different ensemble trajectories of a cell fate induction process. The bifurcation parameter (baseline synthesis rate) dynamics are as detailed in Fig. \ref{fig:BifurcationParam1}, starting at $\eta \approx 0.05$, eventually (a) crossing the tipping threshold ($\eta \approx 0.23$) and (b) approaching, but not crossing it. Time increases along the $z$-axis towards the viewer.}
        \label{Cells3D}
    \end{figure}

    To visualize the tipping behavior of an ensemble of systems under stochastic forcing (combination of B- and N-tipping), we use the representations proposed in Fig. \ref{Inter}, with an added third dimension representing the evolution of the potential landscape in time. We visualize two trajectories of the same dynamical system, one with a bifurcation parameter crossing the tipping threshold (here the baseline synthesis rate $\eta \approx 0.23$), the other with the parameter asymptotically approaching but not crossing it (see Fig. \ref{fig:BifurcationParam1}).
    \enlargethispage{\baselineskip}
    In Fig. \ref{Cells3D}, the observable $x$ corresponds to the phenotype of the cell - with the undifferentiated state on the left and a differentiated state on the right side. Time $t$ increases from the back to the front with two different temporal evolutions of the bifurcation parameter $\eta(t)$ as shown in Fig. \ref{fig:BifurcationParam1}. 

    In the case of B-tipping (Fig. \ref{Cells3D}a), all cells of the ensemble eventually transition to the differentiated state after the system crosses the threshold (i.e., bifurcation point), while some systems start transitioning even before due to stochastic forcing. This can also be seen for the case without B-tipping (Fig. \ref{Cells3D}b), as even though the deterministic threshold is not crossed, noise forces a small number of systems over the potential threshold (forcing cells to commit to a fate). 
    
    It is important to note that these observations and statements can only be made because of the large number $N$ of systems within the ensemble. The individual cells are subject to stochastic influences representing the effect of interactions with each other and the environment, and their exact state is therefore unknown. In the limit of large numbers, however, it is possible to make certain statements about the state density, i.e., the fraction of the ensemble that is in a particular state. The visualization seeks to illustrate this by representing the entire ensemble of cells as a number of small spheres. In this way, the main focus does not lie on the trajectory of a single system, but rather on the dynamics of the whole ensemble.

\subsection{Amazon rainforest}
    One climate component that could potentially exhibit tipping behavior is the Amazon rainforest. Throughout history, the Amazon has acted as a carbon sink, converting carbon dioxide into oxygen through photosynthesis \cite{Hubau2020AsynchronousForests}. Since the 1970s, approximately 17\% of the forest has been eroded \cite{Nobre2021Executive2021} and combined with rising global temperatures and declining precipitation, this has since turned parts of the Amazon into a net source of carbon dioxide \cite{Gatti2021AmazoniaChange}. A further increase in global temperatures would lead to further precipitation decline and a lengthening of the dry season, which could trigger a self-reinforcing feedback loop of drying, forest dieback and savannization \cite{Nobre2021Executive2021}. This risk of a critical transition is exacerbated by deforestation, which could heighten ecosystem vulnerability and shift the tipping point to even lower temperatures \cite{Sampaio2007RegionalExpansion}.

    Here, we model the Amazon rainforest using a simplified version of a TRIFFID model (top-down representation of interactive foliage and flora including dynamics) \cite{Cox2004AmazonianCentury} for a single vegetation type, with modifications and parameter values taken from Ritchie et al. \cite{Ritchie2021}. Its dependent variable $v$ represents the proportion of vegetation in the Amazon ecosystem and is modeled using a logistic equation with an additional decay term 
    \bea
        \frac{\text{d}v}{\text{d}t}=gv(1-v)-\gamma v.
    \eea
    Here, $\gamma=0.2/\text{yr}$ is the disturbance rate representing vegetation mortality and $g$ is the growth rate, a function of the local temperature $T_l$, given by 
    \bea
        g=g_0\left[1-\left(\frac{T_l-T_\text{opt}}{\beta}\right)^2\right]
    \eea
    The maximum growth rate $g_0=2/\text{yr}$ is reached when the local temperature $T_l$ is equal to the optimal growth temperature $T_\text{opt}=28\text{°C}$ and $\beta=10\text{°C}$ determines the half-width of the growth-temperature curve. Finally, there is an additional feedback between the vegetation fraction $v$ and the local temperature; an increase in vegetation cover leads to a lower local temperature $T_l = T_f+\alpha\,(1-v)$. The constant $\alpha=5\text{°C}$ represents the difference in surface temperature between bare soil $(v=0)$ and complete forest cover $(v=1)$, and $T_f$ represents the local temperature if the region were completely covered by vegetation, which depends only on the global temperature and is therefore used as an external forcing parameter. The final differential equation is then given by
    \bea
        \frac{\text{d}v}{\text{d}t}= g_0\, v(1-v)\left [1-\frac{1}{\beta}\biggl(T_f+\alpha(1-v)-T_\text{opt}\biggr)^2\right]-\gamma v.
    \eea
    Starting at $T_f \approx 33.5 \text{°C}$, $T_f$ increases with rising global temperatures, up to a saddle-node bifurcation at $(T_{f,b}, v_b) \approx (34.7\text{°C}, 0.66)$, where the stable equilibrium of a partially vegetated Amazon at $v\approx0.66$ vanishes, causing the system to tip to the remaining stable equilibrium at $v=0$, an Amazon without vegetation cover. Adding noise to the system allows us to model real-world perturbative influences such as weather effects and human intervention. 
    \begin{figure}[H]
        \centering
        \includegraphics[width=0.5\columnwidth]{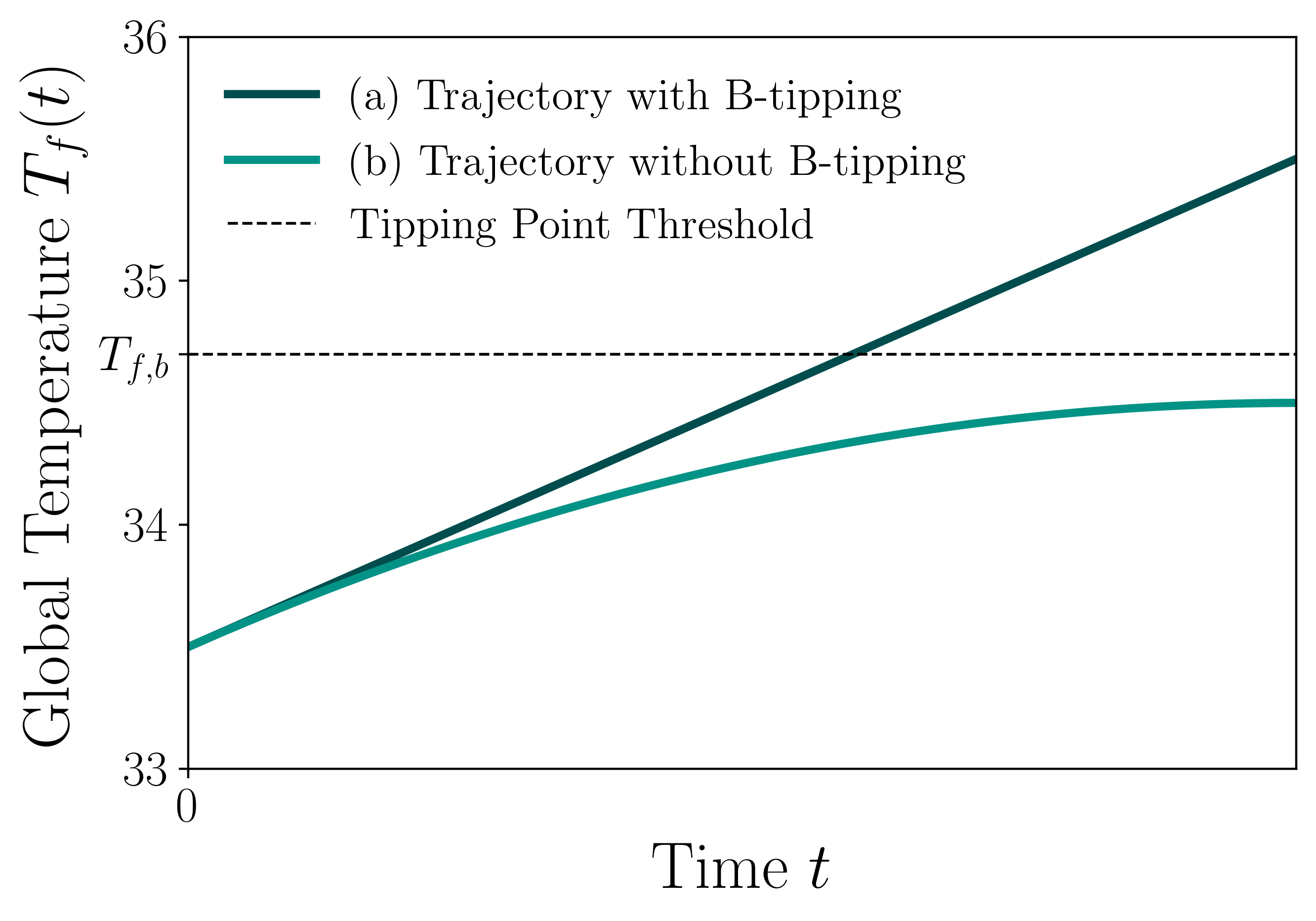}
        \caption{Bifurcation parameter $T_f(t)$ for the two different system trajectories visualized in Fig. \ref{Amazon3D}. Trajectory (a) increases linearly crossing the B-tipping threshold while trajectory (b) approaches but does not cross it.}
        \label{fig:BifurcationParam2}
    \end{figure}
    Again, we present a visualization of two different system dynamics, one that crosses a tipping point boundary and one that does not. In this case, the observable $v$ corresponds to the vegetation fraction of the Amazon. The bifurcation parameter 
    $\eta(t)$ (here $T_f(t)$) is time-dependent similarly to the cell ensemble (see Fig. \ref{fig:BifurcationParam2}).
    
\begin{figure}[!htbp]
        \centering
        \begin{subfigure}[b]{0.49\columnwidth}
            \centering
            \includegraphics[width=\columnwidth]{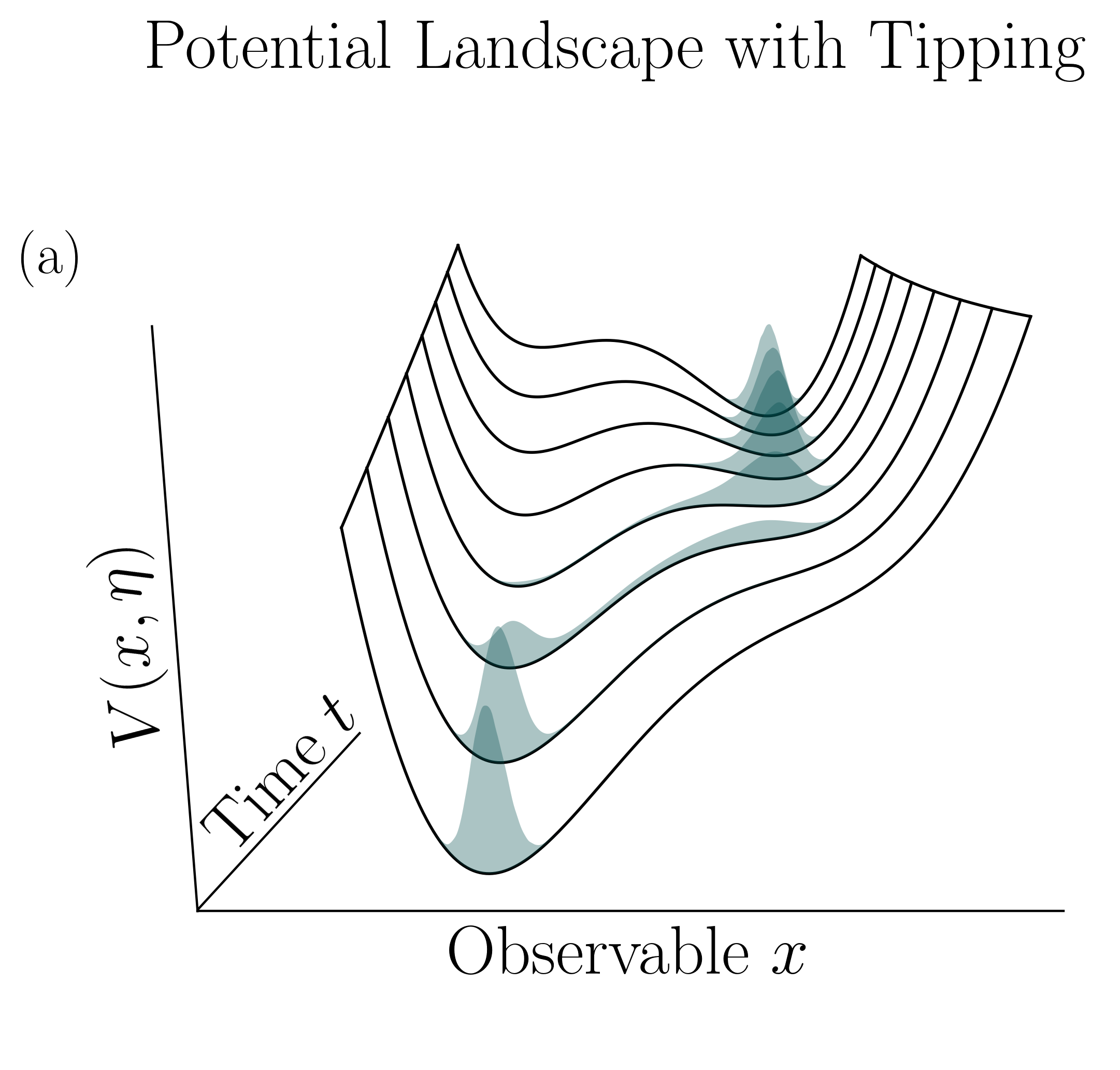}
        \end{subfigure}
        \begin{subfigure}[b]{0.5\columnwidth}
            \centering
            \includegraphics[width=\columnwidth]{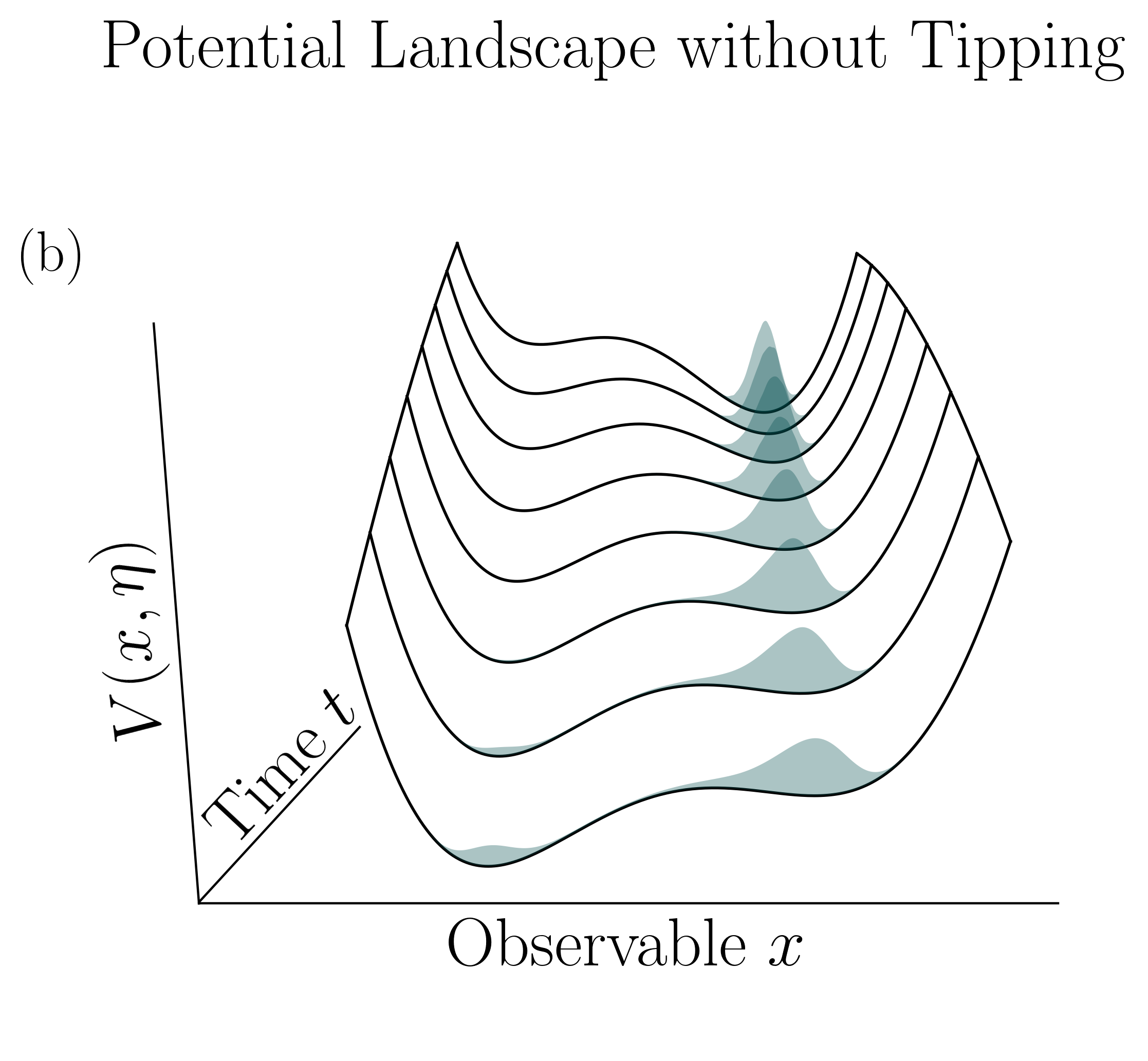}
        \end{subfigure}
        \caption{Potential visualization of two different single system trajectories of a tipping process representing dieback of the Amazon rain forest. The bifurcation parameter dynamics (Indirect measure of global temperature) are as detailed in Fig. \ref{fig:BifurcationParam2}, starting at $33.5 \text{°C}$, eventually (a) crossing the tipping threshold ($T_f \approx 34.7 \text{°C}$) and (b) approaching, but not crossing it. Time increases along the $z$-axis towards the viewer.}
        \label{Amazon3D}
    \end{figure}

     In Fig. \ref{Amazon3D}, the observable corresponds to vegetation fraction of the Amazon - with an unvegetated (died-off) state on the left and a fully vegetated (alive) state on the right side. Time $t$ increases from the back to the front with the bifurcation parameters $\eta(t)$ evolving as detailed in Fig. \ref{fig:BifurcationParam2}. 
     
    Instead of being an ensemble of systems, the Amazon rainforest (as well as all other climate systems and even the Earth respectively) obviously represents only a single system. Therefore, the ensemble interpretation does not apply, and only the evolution of the probability density through time can be considered. In the case of tipping, the system state has a non-zero probability of being in the tipped state even before the tipping point boundary is crossed, corresponding to a dead Amazon without vegetation. This is due to the interaction of B- and N-tipping; a bifurcation-induced flattening of the potential leads to a loss of system resilience, which increases the probability of the noise forcing the system state to transition. After crossing the tipping point, the probability of the system state having transitioned rapidly increases, eventually reaching $p=1$. In the case of no tipping, there is again a small probability of the system being in the tipped state due to the stochastic influence.
    
    Special care must be taken when interpreting this visualization. Climate system dynamics are obviously driven by deterministic physical processes, and stochasticity is introduced only to model internal and external uncertainties and influences of the system. However, it is generally impossible to determine the exact state of the system and its distance to the tipping point, so the exact tipping point boundary can usually only be determined after the system has tipped - an outcome to be avoided in the context of climate dynamics. That is, if the model/visualization shows that the system state has transitioned to a tipped state with some non-zero probability, it is in principle possible to validate whether the system has actually tipped. However, in the context of climate systems, the timescales for tipping to occur after the threshold is reached can vary from years to centuries \cite{McKay2022}. Thus, an increase in the probability of the tipped state within the model signals that the real world system may be approaching a tipping point and that the probability of the system tipping in the future is increasing. Given that there is only one Earth, this seems to be "too risky to bet against" \cite{Lenton2019}. 

\section{Summary and Outlook}
    In this paper, we critically discussed potential pitfalls in visualizations of nonlinear dynamics used in both science communication and research. We discussed the need to distinguish between the author's intended purpose of a visualization and the individual mental model that might be induced in the recipient in several examples, in particular for the "effective potential" (see Fig. \ref{Qpot}) which does not distinguish B- and N-tipping in the visualization. A key contribution of our analysis is the identification of crucial distinctions: the difference between deterministic and stochastic models, and between interpretations of a single system versus an ensemble of systems. Using examples representative of visualizations in popular and academic literature, we have shown that these distinctions are not always made. We therefore propose the visualizations shown in Fig. \ref{Inter} for nonlinear dynamics and discuss examples from molecular biology (cell dynamics, Fig. \ref{Cells3D}, ensemble interpretation) and climate science (Amazon tipping point, Fig. \ref{Amazon3D}, individual system) as explicit applications.
    
    The visualizations we propose are not intended as a final solution, but rather as a theoretically grounded starting point for improved science communication. We acknowledge that the ultimate effectiveness of any visualization can only be validated through empirical research, which, however, requires well-defined and conceptually clear visualizations to test. By explicitly encoding the distinctions between the different cases discussed in \ref{fig:FourField}, we believe our work provides this necessary starting point. The logical next step, therefore, is to employ empirical methods from physics education research and science communication to investigate the mental models that different audiences -- from students to research professionals -- construct from these proposed visualizations. This future research will be essential to test whether our refined representations indeed lead to a more accurate understanding of tipping point dynamics and help to avoid the potential pitfalls identified in this paper.

\bibliographystyle{abbrvnat}
\bibliography{main}

\end{document}